\documentclass{aa}

\usepackage{graphicx}   
\usepackage{amsmath}    
\usepackage{amssymb}    
\usepackage{xcolor}     
\usepackage{soul}   
\usepackage[colorlinks=true,linkcolor=blue,citecolor=blue, urlcolor=blue]{hyperref}
\usepackage{txfonts}
\usepackage{enumitem}
\usepackage{supertabular}
\usepackage{caption}
\usepackage{subcaption}

\usepackage{lipsum}
\usepackage{multirow, booktabs}

\newcommand{\bd}{\begin{displaymath}}
\newcommand{\ed}{\end{displaymath}}
\newcommand{\be}{\begin{equation}}
\newcommand{\ee}{\end{equation}}
\newcommand{\beaa}{\begin{eqnarray*}}
\newcommand{\eeaa}{\end{eqnarray*}}
\newcommand{\bea}{\begin{eqnarray}}
\newcommand{\eea}{\end{eqnarray}}

\newcommand{\sref}[1]{Section~\ref{#1}}

\newcommand{\fref}[1]{Fig.~\ref{#1}}
\newcommand{\fsref}[1]{Figs.~\ref{#1}}
\newcommand{\tref}[1]{Table~\ref{#1}}

\newcommand{\eref}[1]{Eq.~\ref{#1}}

\begin{document} 


\title{Probing compact dark matter objects with microlensing in gravitationally lensed quasars\thanks{Light curves presented in this paper are only available in electronic form at the CDS via anonymous ftp to \href{cdsarc.cds.unistra.fr}{cdsarc.cds.unistra.fr} (130.79.128.5) or via \href{https://cdsarc.cds.unistra.fr/cgi-bin/qcat?J/A+A/}{https://cdsarc.cds.unistra.fr/cgi-bin/qcat?J/A+A/}}}

\titlerunning{Microlensing in lensed quasars}

\author{
P.~Awad\inst{\ref{kapteyn}, \ref{bernoulli}}
\and
J.~H.~H.~Chan\inst{\ref{amnh},\ref{cuny}}
\and
M.~Millon\inst{\ref{epfl},\ref{stanford}}
\and
F.~Courbin\inst{\ref{epfl}}
\and
E.~Paic\inst{\ref{epfl}}
}

\institute{
Kapteyn Astronomical Institute, University of Groningen, PO Box 800, 9700 AV Groningen, The Netherlands
\label{kapteyn}
\newline
\email{p.awad@rug.nl}
\and
Bernoulli Institute for Mathematics, Computer Science and Artificial Intelligence, University of Groningen, 9700AK Groningen, The Netherlands
\label{bernoulli}
\and
Department of Astrophysics, American Museum of Natural History, Central Park West and 79th Street, NY 10024-5192, USA
\label{amnh}
\and
Department of Physics and Astronomy, Lehman College of the CUNY, Bronx, NY 10468, USA
\label{cuny}
\and
Institute of Physics, Laboratory of Astrophysique, \'Ecole Polytechnique F\'ed\'erale de Lausanne (EPFL), Observatoire de Sauverny, 1290 Versoix, Switzerland 
\label{epfl} 
\and 
Kavli Institute for Particle Astrophysics and Cosmology and Department of Physics, Stanford University, Stanford, CA 94305, USA
\label{stanford}
}

\date{\today}

\abstract{
The microlensing signal in the light curves of gravitationally lensed quasars can shed light on the dark matter (DM) composition in their lensing galaxies. Here, we investigate a sample of six lensed quasars from the most recent and best COSMOGRAIL observations: HE~1104$-$1805, HE~0435$-$1223, RX~J1131$-$1231, WFI~2033$-$4723, PG~1115$+$080, and J1206$+$4332, yielding a total of eight microlensing light curves, when combining independent image pairs and typically spanning  ten years. We explore the microlensing signals to determine whether the standard assumptions on the stellar populations are sufficient to account for the amplitudes of the measured signals or whether additional microlenses are needed. We use the most detailed lens models to date from the H0LiCOW/TDCOSMO collaboration to derive the microlensing parameters, such as the convergence, shear, and stellar/dark matter mass fraction at the position of the quasar images. We use these parameters to generate simulated microlensing light curves. Finally, we propose a methodology based on the Kolmogorov-Smirnov test to verify whether the observed microlensing amplitudes in our data are compatible with the most standard scenario, whereby galaxies are composed of stars as compact bodies and smoothly distributed DM. Given our current sample, we show that the standard scenario cannot be rejected, in contrast with previous results by Hawkins~(2020a), claiming that a population of stellar mass primordial black holes (PBHs) is necessary to explain the observed amplitude of the microlensing signal in lensed quasar light curves.
We further estimate the number of microlensing light curves needed to effectively distinguish between the standard scenario with stellar microlensing and a scenario that describes that all the DM contained in galaxies is in the form of compact objects such as PBHs, with a mean mass of $0.2M_{\odot}$. 
We find that about 900 microlensing curves from the Rubin Observatory will be sufficient to discriminate between the two extreme scenarios at a 95\% confidence level. 
}
\keywords{gravitational lensing: micro -- quasars: general -- dark matter}

\maketitle


\section{Introduction} 
\label{sec:intro}

Dark matter (DM) is a hypothetical form of matter that is thought to account for $\sim80\%$ of mass in a galaxy. Although we generally see a broad consensus regarding the gravitational effects of DM, its nature remains largely unknown. Overall, DM is theorized to either take the form of a particle \citep[for a review, see e.g.,][]{BoyarskyEtal19} that is smoothly distributed or the form of compact objects such as massive compact halo objects \citep[MACHOs;][]{AlcockEtal92, AubourgEtal94}. The search for the true form that would accurately describe the nature of DM is essential, as this query lies at the foundation of efforts to build cosmological models of the Universe.

One of the phenomena that has been used as a means to study the composition of galaxies is microlensing. In strongly lensed quasars, the gravitational lensing effect produces multiple images of the quasar source. Subsequently, microlensing is induced by compact objects such as stars in the lensing galaxies \citep{Chang&Refsdal79,Gott81,IrwinEtal89}: the light from the source passing by these compact objects is further split to produce "micro-images" that are separated from each other by a few $\mu$-arcseconds \citep[see a review in][]{Wambsganss06}. These resulting micro-images cannot be resolved, but they can still be detected through the anomalous fluxes in the strongly lensed images. Microlensing induces variations in the quasar brightness on timescales of several months to years \citep{Mosquera&Kochanek11}, enabling us to study the properties of lensing galaxies, such as their dark matter fractions \citep{Schechter&Wambsganss04}, stellar mass functions \citep{Jimenez-Vicente&Mediavilla19}, and the properties of quasar accretion disks \citep{MorganEtal08, EigenbrodEtal08, MorganEtal10, Jimenez-VicenteEtal19, CornachioneEtal20}.  

One aspect that has been generally assumed for the origin of microlensing signals relates to the stellar populations inhabiting the halos of the lensing galaxies \citep{Schild90,FalcoEtal91,Kundic&Wambsganss93}. However, in observed light curves of strongly lensed quasars, the stellar distributions are not always sufficient for explaining microlensing flux variations \citep{MediavillaEtal09,PooleyEtal12}. This resulting issue is, in fact, tied to the nature of DM particles. It has been argued that compact bodies in the form of primordial black holes (PBHs) can offer a better explanation for the microlensing events in the light curves of lensed quasars than stellar populations \citep{H20,H20b}. These works have been reinforced with the recent detection of gravitational waves from a black hole merger \citep{AbbottEtal16}, which lends support to the idea that PBHs may make up a significant fraction of the DM in galaxy halos \citep{BirdEtal16,SasakiEtal16,ByrnesEtal18}. In other works, galactic microlensing has been used to set the limits of PBH abundances for different ranges of masses \citep[e.g.,][]{AlcockEtal01,TisserandEtal07,WyrzykowskiEtal11,Blaineau22}. Accordingly, microlensing is a prominent technique for the exploration of PBH abundance and possibly provides an insight into the nature of DM. 

In this work, we build on what has been presented in \cite{H20,H20b}, with the intention of exploring whether stars in lensing galaxies of lensed quasars can indeed account for their microlensing signals. If the stellar distributions are proven to be insufficient, then a fraction of another form of compact bodies is needed to explain the variability in the observed microlensing signals of lensed quasars. Compared to the study of \cite{H20,H20b}, here we take a larger sample of lensed quasar systems that contains light curves with long baselines measured by COSMOGRAIL\footnote{\href{http://www.cosmograil.org}{http://www.cosmograil.org}} \citep{CourbinEtal05,MillonEtal20b}. To explore the effect of stellar distributions on the microlensing curves of these lens systems, we utilized the latest lensing model parameters provided by the H0LiCOW\footnote{\href{https://shsuyu.github.io/H0LiCOW/site/}{http://h0licow.org}}/TDCOSMO\footnote{\href{http://www.tdcosmo.org}{http://www.tdcosmo.org}} collaboration \citep{SuyuEtal17,MillonEtal20a}. Using the simulated microlensing signals, we compared the amplitude to that seen in their observational counterparts. 

The paper is structured as follows. In \sref{sec:data}, we describe the data used in this work, including the chosen sample of lensed quasars, followed by the extracted microlensing signals in observational and simulated light curves. The description of the methodology for our statistical analysis is then provided in \sref{sec:method}. We present our results in \sref{sec:results} with the proposal for future applications using data from the Rubin observatory in \sref{sec:prediction}. In \sref{sec:discussion}, we discuss the assumptions in our analysis as well as the obtained results. Our conclusions are given in \sref{sec:conclusion}. Throughout this work, we use a flat-$\Lambda$CDM cosmology, with $H_0=70~{\rm km~s^{-1}~Mpc^{-1}}$, $\Omega_m=0.3,$ and $\Omega_\Lambda=0.7$.
 

\section{Data}
\label{sec:data}

Our goal is to investigate the nature of the compact objects in lensing galaxies that are responsible for the presence of microlensing events in the light curves of lensed quasars. The sample of lensed quasar systems is described in \sref{subsec:sample}, followed by the methods used to extract the microlensing signals of the observational data in \sref{subsec:observations}. We then describe in \sref{subsec:simulations} the simulation methodology and setup that we use for our comparison with the observational data.

\subsection{Lensed quasar sample}
\label{subsec:sample}

To build a sample of lensed quasars, we looked through the most recent light curve measurements of COSMOGRAIL \citep{MillonEtal20b} and selected the systems that have light curve measurements for longer than five years, as well as those with lensing parameters modeled by the H0LiCOW and TDCOSMO collaborations. In \tref{table:MacroModels}, we list the lensing parameters at the position of each lensed quasar image, including the total convergence, $\kappa$, the stellar convergence, $\kappa_*$, and the total shear, $\gamma$. Based on this criteria, our sample consists of five lens systems, namely: HE~0435$-$1223, RX~J1131$-$1231, WFI~2033$-$4723, PG~1115$+$080, and J1206$+$4332 (hereafter, HE0435, RXJ1131, WFI2033, PG1115, and J1206, respectively). To this sample, we added the system HE~1104$-$1805 (hereafter, HE1104) to allow for a comparison with the work of \citet{H20}. The derivation of the lensing parameters for the latter system is explained in \sref{subsec:simulations}, which follows similar procedures as all the other systems modeled by TDCOSMO.

\begin{table*}[]
  \caption{Properties of the chosen lensed quasar systems along with the references for their lensing parameters.}
  \centering
  \begin{tabular}{p{1.25cm} p{0.45cm} p{0.45cm} p{1.75cm} p{2cm} p{0.75cm} p{0.9cm} p{0.9cm} p{0.9cm} p{1.5cm} p{2.75cm}}

    \toprule

    {\bfseries System} & {$z_s$} &  {$z_l$} & {$R_E$ [$10^{16}$cm]} & {$R_{1/2}$ [$10^{16}$cm]} & {\bfseries Image} & {\bfseries $\quad \kappa$} & {\bfseries $\quad \kappa_*$} & {\bfseries $\quad \gamma$ } & {$\mathcal{M}$ [mag]} & { \bfseries Reference}\\
    \midrule\midrule[.1em]

    \multirow{2.5}{1.25cm}{HE1104}
     &\multirow{2.5}{0.5cm}{2.32} 
     &\multirow{2.5}{0.5cm}{0.73} 
     &\multirow{2.5}{1.75cm}{1.933}
     &\multirow{2.5}{1.75cm}{0.668}
      & {\ttfamily \quad A}
      &  0.648
      &  0.125
      &  0.605
      &  $-1.540$
      &\multirow{2.5}{2.75cm}{The current work}\\
    \cmidrule(lr){6-10}
      &
      &
      &
      &
      & {\ttfamily \quad B}
      & 0.332
      & 0.019
      & 0.290
      & $-1.103$\\

    \midrule[.1em] 

    \multirow{5.5}{2cm}{HE0435}
      &\multirow{5.5}{1cm}{1.69}
      &\multirow{5.5}{1cm}{0.46}
      &\multirow{5.5}{1cm}{2.406}
      &\multirow{5.5}{1cm}{0.885}
      & {\ttfamily \quad A}
      & 0.473
      & 0.164 
      & 0.358
      & $-2.063$
      &\\
    \cmidrule(lr){6-10}
      &
      &
      &
      &
      & {\ttfamily \quad B}
      & 0.630
      & 0.227 
      & 0.540
      & $-2.026$
      & \multirow{2.5}{3.25cm}{\citet{ChenEtal19}} \\
    \cmidrule(lr){6-10}
      &
      &
      &
      &
      & {\ttfamily \quad C}
      & 0.494
      & 0.165
      & 0.327
      & $-2.066$
      &\\

    \cmidrule(lr){6-10}
      &
      &
      &
      &
      & {\ttfamily \quad D}
      & 0.686
      & 0.260
      & 0.575
      & $-1.586$ 
      &\\
      
    \midrule[.1em] 

    \multirow{5.5}{2cm}{RXJ1131}
      &\multirow{5.5}{1cm}{0.66}
      &\multirow{5.5}{1cm}{0.29}
      &\multirow{5.5}{1cm}{2.081}
      &\multirow{5.5}{1cm}{1.685}
      & {\ttfamily \quad A}
      & 0.526
      & 0.226
      & 0.410
      & $-3.118$
      & \\
    \cmidrule(lr){6-10}
      &
      &
      &
      &
      & {\ttfamily \quad B}
      & 0.459
      & 0.199
      & 0.412
      & $-1.926$
      & \multirow{2.5}{3.25cm}{\citet{ChenEtal19}}\\
    \cmidrule(lr){6-10}
      &
      &
      &
      &
      & {\ttfamily \quad C}
      & 0.487
      & 0.201
      & 0.306
      & $-2.276$
      &\\
    \cmidrule(lr){6-10}
      &
      &
      &
      &
      & {\ttfamily \quad D}
      & 0.894
      & 0.519
      & 0.807
      & $-0.493$
      &\\

    \midrule[.1em] 
    
    \multirow{2.5}{2cm}{WFI2033}
      &\multirow{2.5}{1cm}{1.66}
      &\multirow{2.5}{1cm}{0.66}
      &\multirow{2.5}{1cm}{1.939}
      &\multirow{2.5}{1cm}{0.898}
      & {\ttfamily \quad B}
      & 0.445
      & 0.145 
      & 0.208
      & $-1.443$
      & \multirow{2.5}{3.25cm}{\citet{RusuEtal20}}\\
    \cmidrule(lr){6-10}
      &
      &
      &
      &
      & {\ttfamily \quad C}
      & 0.792
      & 0.396
      & 0.538
      & $-1.522$
      &\\

    \midrule[.1em] 

    \multirow{2.5}{2cm}{PG1115}
      &\multirow{2.5}{1cm}{1.72}
      &\multirow{2.5}{1cm}{0.31}
      &\multirow{2.5}{1cm}{2.959}
      &\multirow{2.5}{1cm}{0.872}
      & {\ttfamily \quad B}
      & 0.502
      & 0.166 
      & 0.811
      & $-0.968$
      & \multirow{2.5}{3.25cm}{\citet{ChenEtal19}}\\
    \cmidrule(lr){6-10}
      &
      &
      &
      &
      & {\ttfamily \quad C}
      & 0.356
      & 0.072
      & 0.315
      & $-1.252$
      &\\
    \midrule[.1em] 
    
    \multirow{2.5}{2cm}{J1206}
      &\multirow{2.5}{1cm}{1.79}
      &\multirow{2.5}{1cm}{0.85}
      &\multirow{2.5}{1cm}{1.671}
      &\multirow{2.5}{1cm}{0.843}
      & {\ttfamily \quad A}
      & 0.656
      & 0.095
      & 0.685
      & $-1.137$
      & \multirow{2.5}{3.25cm}{\citet{BirrerEtal19}}\\
    \cmidrule(lr){6-10}
      &
      &
      &
      &
      & {\ttfamily \quad B}
      & 0.401
      & 0.020
      & 0.364
      & $-1.613$
      &\\
      
    \midrule\midrule[.1em] 

  \end{tabular}
  \tablefoot{From left to right, we give: the name of the system in the sample, source redshift ,$z_s$, lens redshift, $z_l$, Einstein radius, $R_E$, half-light radius of the lensed quasar disk, $R_{1/2}$, as computed in \eref{eq:wavelengthScale}, chosen images for each system, total convergence $\kappa$, stellar convergence, $\kappa_*$, total shear, $\gamma$, macro-magnification, $\mathcal{M,}$ using \eref{eq:macromag}, and the reference for the lens models. The Einstein radius of the microlenses $R_E$ is calculated following \eref{eq:EinRadius}, with $\langle M_*\rangle=0.2 M_\odot$ and assuming a Salpeter IMF.}
  \label{table:MacroModels}
\end{table*}

\begin{figure*}
\centering
\includegraphics[scale=0.65]{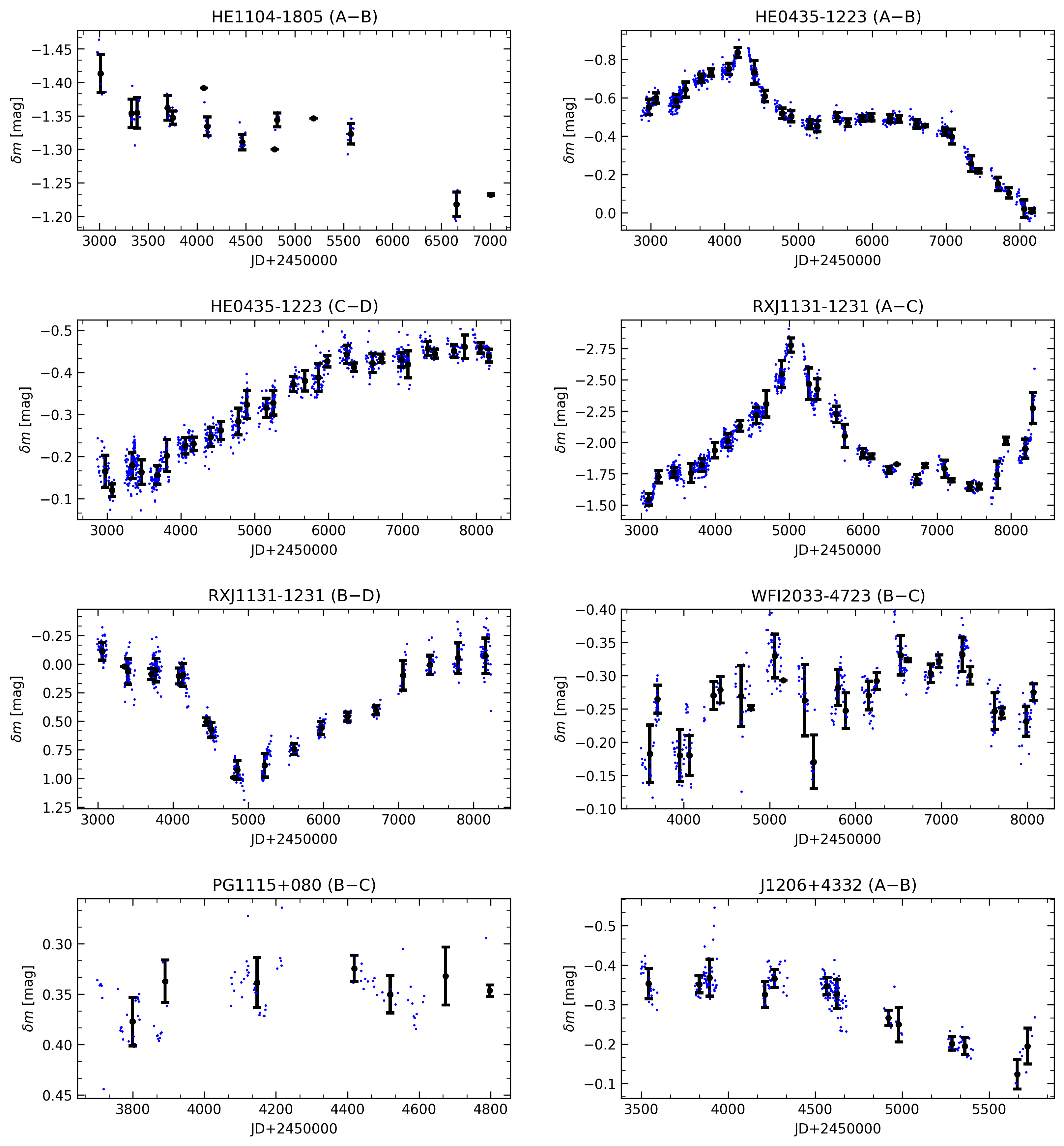}
\caption{
Microlensing curves obtained from the COSMOGRAIL observations. The macro-magnifications computed from the lens models have been subtracted. The curves are half-yearly binned with the weighted mean and the standard deviation as the error bars, shown as black points. The original unaveraged measurements are shown in blue. 
}

\label{fig:MicrolensingLC}
\end{figure*}

The observed light curves contain not only the signal of microlensing in lensing galaxies, but also the intrinsic variability of quasar sources. In order to extract the microlensing signals, we removed the intrinsic variation of the quasar by shifting the curves of a pair of lensed images with their time delay and subtracting them (see \sref{subsec:observations} for  more). We refer to the resulting difference light curve the "microlensing curve" in the rest of this paper. We used only independent pairs of light curves from each system to avoid accounting twice for the same microlensing event, which would then appear in all microlensing curves computed relative to a given lensed image. In the cases of doubly lensed quasars (doubles): HE1104 and J1206, there is only one pair available, so we obtained only one microlensing curve. The quadruply lensed quasars (quads) have two independent pairs of images, thus providing us with two independent microlensing curves. In the case of WFI2033 and PG1115, which are both quads in fold configuration, the two closest images are not resolved in the monitoring data. Their image As are actually a blend of two bright images, which makes the microlensing signals of the unresolved image intractable. Thus, we only considered images B and C for these two systems. Finally, we first considered images A and B for HE0435, along with images A and C for RXJ1131, in order to match the choice of \citet{H20}. Since there are an additional two images, we included the image pairs of C and D for HE0435 and of B and D for RXJ1131, respectively. In total, our sample consists of eight independent pairs of lensed images that made up the basis of our subsequent analysis. The considered image pairs are listed in \tref{table:pvalues} and their microlensing curves are shown in \fref{fig:MicrolensingLC}.

\begin{center}
  \begin{table}[h!]
  \caption{Microlensing amplitudes and p-values, $p$ (obtained in this work) and $p_{\rm H20}$ obtained by \cite{H20}.} 
  \setlength{\tabcolsep}{8pt} 
  \renewcommand{\arraystretch}{1.5} 
  \centering
  \begin{tabular}{p{1.5cm} p{1cm} p{1cm} p{1cm} p{1cm}} 
   \hline
   System & pair & $\Delta m_{\rm obs}$ [mag] & $p$ & $p_{\rm H20}$\\ 
   \hline\hline
   HE1104 & A -- B & 0.195 & 0.762 & 0.046
   \\ 
   \hline
   HE0435 & A -- B & 0.826 & 0.377 &  0.050
   \\
   \hline
   HE0435 & C -- D & 0.340 & 0.378 &  --
   \\
   \hline
   RXJ1131 & A -- C & 1.231 & 0.190 &  0.000
   \\
   \hline
   RXJ1131 & B -- D & 1.103 & 0.143 & -- 
   \\
   \hline
   WFI2033 & B -- C & 0.161 & 0.787 & 0.131
   \\ 
   \hline
   PG1115 & B -- C & 0.053 & 0.610 & --
   \\ 
   \hline
   J1206 & A -- B & 0.245 & 0.283 & --
   \\ 
   \hline
   
  \end{tabular}
  \label{table:pvalues}
  \end{table}
\end{center}

\begin{figure*}[h!]  
\centering
\begin{subfigure}{0.78\textwidth}
\includegraphics[width=\linewidth]{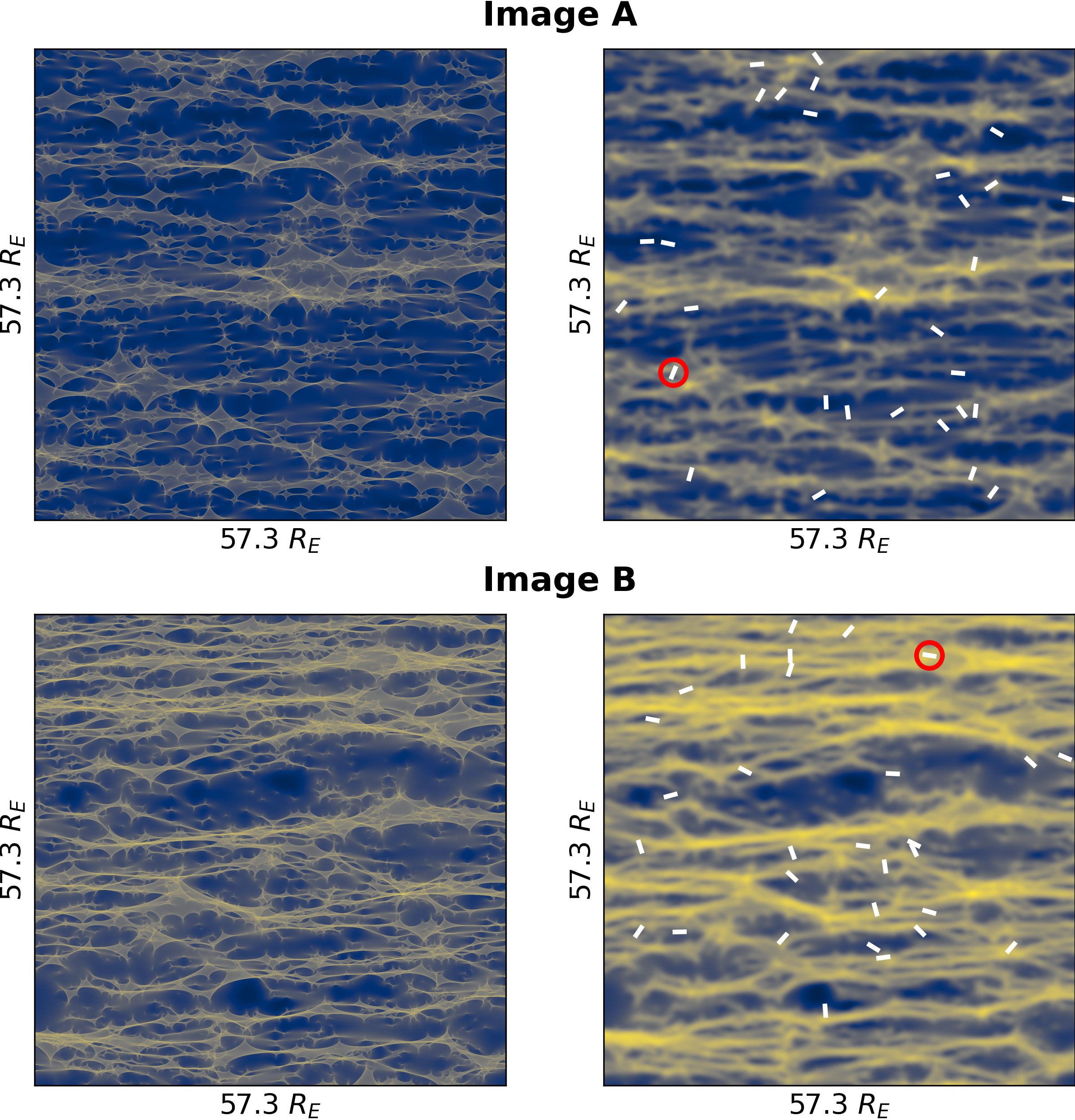}
\end{subfigure}

\bigskip

\begin{subfigure}{0.8\textwidth}
\includegraphics[width=\linewidth]{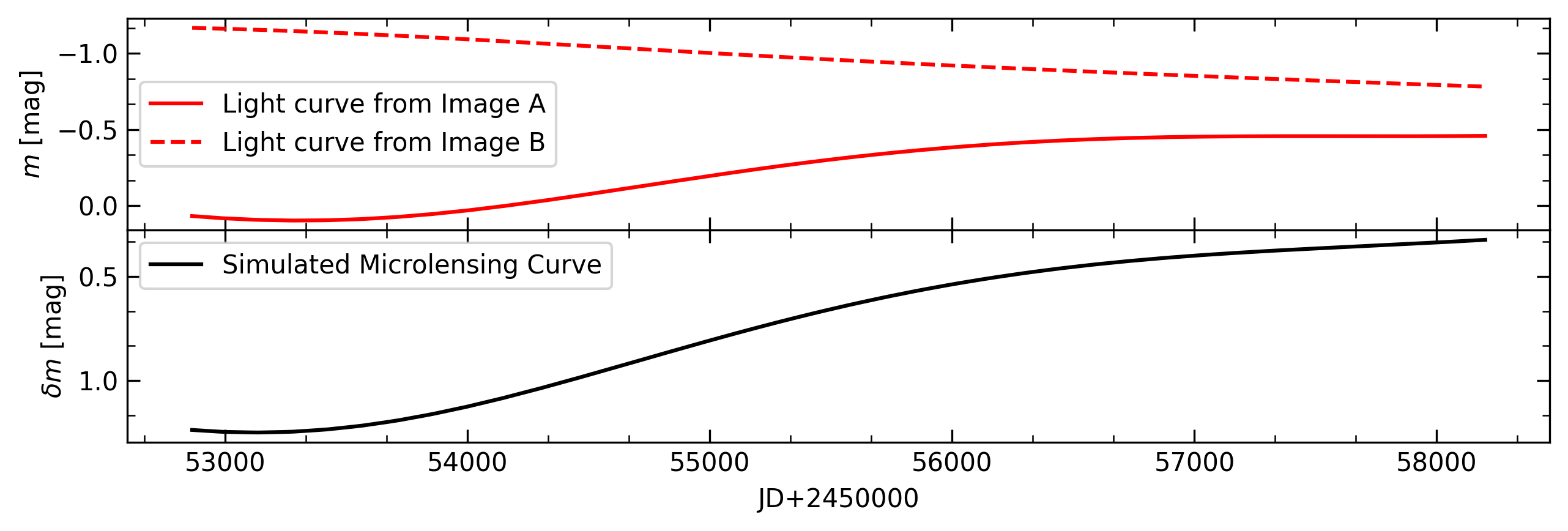}
\end{subfigure}
\caption{
Microlensing simulation for HE~0435$-$1223. Magnification maps of images A (top left) and B (middle left) are generated using microlenses with $\langle M_*\rangle=0.2$ following the Salpeter initial mass function within the mass range of $0.06M_{\odot}$ and $6.46M_{\odot}$. The convolved maps of images A (top right) and B (middle right) are obtained with a Gaussian kernel for the source half-light radius as listed in \tref{table:MacroModels}. Each map has $57.3~\times~R_E$ on a side (or 2000 pixels). A few trajectories (white lines) are randomly drawn with a transverse velocity of 600~km/s and a length equal to the observational duration (5338 days). A pair of simulated light curves (red circles) is 
shown on the bottom panel. The corresponding difference light curve $\delta m$ yields the microlensing amplitude $\Delta m\approx 0.75$~mag, following our definition in Sect.~\ref{sec:method}. 
} 
\label{fig:MagMaps}
\end{figure*}

\subsection{Microlensing light curves from observations}
\label{subsec:observations}

We can obtain microlensing signals by shifting a pair of observed light curves with its respective time delay. For example, given a pair of lensed images labeled $i$ and $j$, assuming that image $j$ occurs with a time delay, $\Delta t$, we can describe the light curve for each image as a function of time, $t,$ in the unit of magnitude:
\begin{equation}
    \begin{aligned}
        S_{i}(t) &= I(t) + \mathcal{M}_{i} + m_{i}(t),\\
        S_{j}(t) &= I(t-\Delta t) + \mathcal{M}_{j} + m_{j}(t-\Delta t),\\
    \end{aligned}
\label{eq:signal}    
\end{equation}
where $I(t)$ is the quasar intrinsic variation, $m(t)$ is the microlensing variations, and $\mathcal{M}$ is the macro-magnification induced by the lensing setup. We can express the macro-magnification factor in the unit of magnitude as the inverse of the Jacobian of the lensing matrix \citep{SchneiderEtal92}:
\begin{equation}
    \mathcal{M} = -2.5 \log_{10}\Bigg(\Big\lvert{\frac{1}{(1-\kappa)^2 - \gamma^2}} \Big\rvert \Bigg).
    \label{eq:macromag}
\end{equation}

After shifting the light curve of image $j$ by the time delay, $\Delta t$, we subtract the induced macro-magnification, $\mathcal{M,}$ in each image and compute the difference between the signals, $S_{i}$ and $S_{j}$. Since the intrinsic signal of the source is the same in both lensed images, the corresponding term will cancel out upon taking the difference between the two signals of the pair, so that we are able to retain the difference between the microlensing contributions from each image. In other words, from \eref{eq:signal}, we can derive a microlensing curve, $\delta m,$ using a pair of lensed images:
\begin{equation}
\begin{aligned}
     \delta m (t) &= S_{i}(t) - S_{j}(t + \Delta t) - \left(\mathcal{M}_{i} - \mathcal{M}_{j}\right)\\
                  &= m_{i}(t) - m_{j}(t).\\
\end{aligned}
\label{eq:difference_obs}
\end{equation}
We employed \texttt{PyCS3}\footnote{\url{https://cosmograil.gitlab.io/PyCS3/}} \citep{TewesEtal13,MillonEtal20c}, a curve-shifting python package, to fit, shift, and subtract the light curves. In this work, we are only interested in the long-term microlensing variations attributed to stars passing in front of the quasar images and modulating the microlensing magnification. These variations occur on a time scale of months to years. Hence, we re-binned our data half-yearly to reduce the photometric noise within the fluctuating signals. We calculated the weighted mean of each bin, taking the weights on both axes as the inverse of the error on the magnitude measurements for each point in a given bin. The error for each bin represents the standard deviation of all the photometric measurements within that bin. Our original and binned observational microlensing curves are illustrated in \fref{fig:MicrolensingLC} by the blue and black data points, respectively.

\subsection{Microlensing light curves from simulations}
\label{subsec:simulations}

In order to test if stars in lensing galaxies can explain the observed microlensing signal, we simulated light curves drawn randomly from microlensing magnification maps for each quasar image. We generated the magnification maps using GPU-D, an inverse ray-shooting code on the GPU \citep{Vernardos&Fluke13} that efficiently computes  networks of lensing microcaustics in the source plane. This requires an estimate of the total projected mass, $\kappa$, shear, $\gamma$, and projected fraction of mass under the stellar form, $\kappa_*$, at the position of the quasar images. 

For the quadruply imaged quasars, we considered a composite mass model for the lens to infer $\kappa$, $\kappa_*$, and $\gamma$, following \citet{Suyu2014} and \cite{ChenEtal19}. This is also the methodology adopted by the H0LiCOW \citep{WongEtal20} and TDCOSMO collaborations \citep[e.g.,][]{MillonEtal20a}. In these works, the lens mass has two components. The first represents the stellar mass, namely, a fit to the 2D projected light distribution of the lens, scaled by a radially constant mass-to-light ratio. The second is a standard Navarro-Frenk-White (NFW) profile \citep{NFW96}. The stellar and dark mass components are fitted jointly and the mass-to-light ratio of the stellar component is a free parameter during this fit. The resulting values for the models are given in \tref{table:MacroModels}, along with the reference of the papers that specifically studied these lenses. 

For HE1104, which only has two lensed images, a composite model has too many degrees of freedom. We instead used a simpler power-law
elliptical mass distribution with shear (PEMD + shear), as implemented in the Lenstronomy package \citep{Birrer2018}. In this model, the stellar and dark mass components cannot be treated separately. We therefore followed the approach of \citet{auger2009}, who modeled a large sample of (lensing) early-type galaxies (ETG) with power-law models and found that the mean fraction of stellar mass within half of the effective radius, that is, where the lensed images fall, is $f_* = 0.7$. To find $\kappa_*$ we first integrate the lens light in our HST images in the F160W band, in the same aperture as \citet{auger2009}. We then assume that $f_* = 0.7$ is constant across the galaxy and integrate the convergence of the mass model in the same aperture as for the light and compute the normalization factor to apply on the lens light at the image position to obtain $\kappa_*$. The values for HE1104 are listed in \tref{table:MacroModels} along with the values for the other systems.

It is also possible to calculate the lens models by analyzing the microlensing effects of flux measurements for individual lensed quasars \citep{MediavillaEtal09, PooleyEtal12, JimenezVicenteEtal15, EstebanEtal22}. This method, however, can be unreliable as it depends on several assumptions that can lead to underestimating the stellar mass fraction. Some of these assumptions include presuming the source to be infinitely compact and attributing flux ratio anomalies entirely to microlensing. Such anomalies, however, can also arise from millilensing by substructures in the main deflector or along the line of sight. Correcting for these assumptions would lead to higher estimates of stellar fractions to explain measured flux ratios. We therefore rely on the lens modeling performed by the TDCOSMO collaboration for detailed modeling of each lensed system.

The mass under the form of stars is distributed following a Salpeter initial mass function (IMF), with the mean mass of $\langle M_* \rangle = 0.2M_{\odot}$ and the mass ratio between the heaviest and lightest microlenses fixed to be $100$ in the mass range $0.06M_{\odot}$ and $6.46M_{\odot}$ \citep{ChanEtal21}. Other IMFs have not been explored since they do not produce a noteworthy effect on the results \citep{KochanekEtal07, WyitheEtal00}.

The size of a magnification map is scaled to the Einstein radius of microlenses, $R_E$, on the source plane, described as:
\begin{equation}
    R_E = D_{\mathrm{S}} \times \sqrt{\frac{4G\langle M_*\rangle}{c^2} \frac{D_{\rm LS}}{D_{\rm L}D_{\rm S}}},
\label{eq:EinRadius}
\end{equation}
which depends on the angular diameter distances from the observer to the lens, $D_{\rm L}$, from the observer to the source, $D_{\rm S}$, and between the lens and the source, $D_{\rm LS}$. The Einstein radius for each system is provided in the fourth column of \tref{table:MacroModels}, assuming $\langle M_* \rangle=0.2 M_\odot$ and the source and lens redshifts listed in the second and third columns of the same table.

In \fref{fig:MagMaps}, we show the magnification maps of images A and B of HE0435 as an example. The observed flux at a particular source position can be computed by convolving the quasar's accretion disk's light profile with the micro-caustics pattern. This, of course, depends on the size of the accretion disk. In this work, we estimate the accretion disk sizes of $15$ quasars with continuum reverberation mapping from \citet{MuddEtal18}, which avoids any circular arguments caused by choosing disk sizes measured with previous assumptions on the stellar populations of the lensing galaxies \citep[e.g.,][]{MorganEtal10,MorganEtal18,CornachioneEtal20}. We further discuss the choice of disk sizes in \sref{sec:discussion}. 

In order to estimate the disk size for each lens system, we started from the estimates of \cite{MuddEtal18} for the mean black hole mass ($5.5\times10^8~M_\odot$) and for the flux-weighted mean disk size at the rest wavelength $\lambda=2500~\AA$ ($7\times10^{15}$~cm). We then scaled the disk sizes with the mean black hole mass of our sample, $\overline{M}_{\rm BH}=8.35\times10^8~M_\odot$, and we corrected for the redshift of each individual lensed quasars to obtain an estimate of the disk size at the rest-frame wavelength $\lambda_{\rm rest}=\lambda_{\rm obs}/(1+z_s)$. The COSMOGRAIL observations are taken in the R band, which corresponds to $\lambda_{\rm obs} = 6500~\AA$. We chose to scale the disk sizes by the mean -- and not by the individual black hole mass, since the black hole masses of lensed quasars 
may not all be well constrained. We therefore used the mean mass of our sample to avoid any biasing that could occur in the case of one or more of the mass estimates being highly inaccurate, although this choice does not affect our analysis significantly. The half-light radii of the disks can be expressed as:
\begin{equation}
R_{1/2} =  7\times10^{15}~{\rm cm}\ \Big(\frac{\lambda_{\rm rest}}{2500\AA} \Big)^{4/3}\Big(\frac{\overline{M}_{\rm BH}}{5.5\times10^8~M_\odot} \Big)^{2/3},
\label{eq:wavelengthScale}
\end{equation}
and they are listed in \tref{table:MacroModels} for each of the six systems of our sample. We note that this scaling relation follows the thin-disk model \citep{Shakura&Sunyayev73}, which is supported by microlensing observations \citep[e.g.,][]{Eigenbrod2008}. The mean half-light radius of our sample is then $\overline{R_{1/2}}\sim10^{16}$~cm, which corresponds to about four light-days and is comparable with the choice of \citet{H20}.
The disk light distribution is also assumed to be Gaussian to match the light profile used in \citet{H20}. The choice of disk models has been reported to have only a minor effect on the expected amplitude of the microlensing signal \citep{MortonsonEtal05}. We generated the magnification maps with the size of $72~\times~R_E$ in $2512$~pixels on each side and the disk light distribution with the size of $14.7~\times~R_E$ in $512$~pixels. As the convolution can slightly reduce the map size, the final effective map size is $57.3~\times~R_E$, with a resolution of $2000$~pixels on a side.

After we obtained the convolved maps, we drew random trajectories on each map to simulate the light curves. In this work, we assume (conventionally) a transverse velocity of $600$~km/s for the microlenses. More precise estimates are available \citep{Neira2020} but our choice of velocity also matches that of \citet{H20}. The time duration of each simulated light curve is the same as the observed one (shown in \fref{fig:MicrolensingLC}). Since the observed microlensing signal is the outcome of the difference between a pair of lensed images, we also computed the difference between the simulated light curves from two maps. Given the light curves of a lensed image pair in flux units $F_{i}(t)$ and $F_{j}(t)$, the simulated difference curve in magnitude can be expressed as: 
\begin{equation}
\begin{aligned}
     \delta m (t) &= -2.5 \log_{10}\Bigg(\frac{F_{i}(t)}{F_{j}(t)} \Bigg) - \left(\mathcal{M}_{i} - \mathcal{M}_{j}\right)\\
                  &= m_{i}(t) - m_{j}(t),\\
\end{aligned}
\label{eq:difference_sim}
\end{equation}
where $\mathcal{M}$ is the macro-magnification obtained from the lens macro model, equal to that adopted for the observed light curves, as in \eref{eq:difference_obs}. We illustrate a few simulated light curves, $m(t)$, using tracks in the microlensing caustics. Examples of such tracks are indicated as white lines in \fref{fig:MagMaps}. We note that $m=0$ corresponds to no microlensing. An example of a pair of curves (red circles) is highlighted on the bottom panel of the figure.

\section{Method}
\label{sec:method}

\begin{figure*}[h]
        \includegraphics[width=\textwidth]{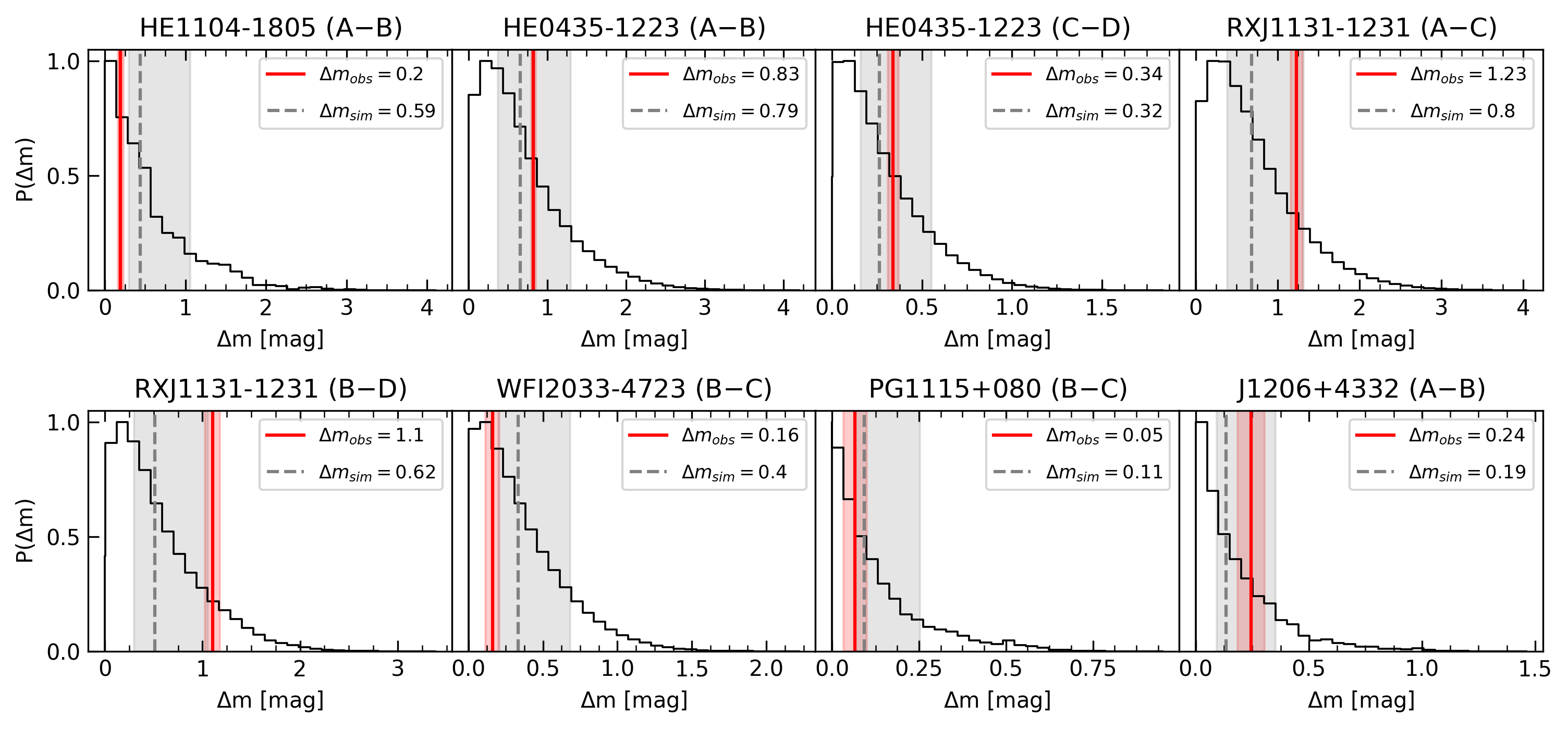}
    \caption{Probability distributions of magnification amplitude, $\Delta m$. The black histogram is generated from $\Delta m_{\rm sim}$ in the simulated light curves, with the median (dashed line), along with the 84th and 16th percentile indicated as grey shaded areas. The red vertical lines, $\Delta m_{\rm obs}$, are measured in the observed light curves and given in \tref{table:pvalues}. We also display the quadratic sum of the photometric uncertainties as a red shaded line.}
    \label{fig:InitialModels}
\end{figure*}

We went on to compare the amplitude of the microlensing signals in observations and simulations. We adopted the difference between the maximum and the minimum magnitudes as the definition for the amplitude of each curve:
\begin{equation}
    \Delta m = \max(\delta m (t)) - \min(\delta m (t)),
\label{ourAmp}
\end{equation}
which follows the choice of \cite{H20b}. Importantly, this definition is different from \cite{H20}, where $\Delta m$ is defined as the maximum deviation from the difference in macro-magnification ($\mathcal{M}_{i}-\mathcal{M}_{j}$), dubbed "zero points." The "zero point" can be estimated from the lens modeling or from the flux ratios if the observing wavelength corresponds to an emission region that is sufficiently large not to be affected by microlensing (e.g., the broad line regions or the extended radio emission regions). This definition, however, is prone to significant biases, as the amplitude is directly affected by the level of accuracy of either the lens models or the observational measurements. Instead, our definition in \eref{ourAmp} is purely empirical. It does not rely on the precision of the lens models since its respective zero points are canceled out when calculating the difference between the maximum and minimum of a difference curve. This choice is therefore more robust since it allows us to study the microlensing variations over several years without using any information about the absolute amplitude of the microlensing magnification. We list in \tref{table:pvalues} our measurements of $\Delta m$ on the observed difference light curves in \fref{fig:MicrolensingLC}.

The amplitude, $\Delta m,$ from the simulation was calculated in the exact same way as for the observational curves. We indicate the amplitudes from simulations as $\Delta m_{\rm sim}$ and those from observations as $\Delta m_{\rm obs}$. For each image pair, we drew $10^3$ light curves $m(t)$ randomly oriented on the corresponding magnification maps and subtracted all pairs of light curves using \eref{eq:difference_sim}, resulting in $10^6$ simulated difference curves $\delta m(t)$. The probability distribution of $\Delta m$ from simulated light curves, $P(\Delta m)$, is shown in \fref{fig:InitialModels} for each system. The observed amplitudes, $\Delta m_{\rm obs}$, are labeled as red lines with observational errors corresponding to the sum in quadrature of the the maximum and the minimum photometric uncertainties for each observational difference curve. By counting the frequency with which the amplitude of the simulated light curve, $\Delta m_{\rm sim}$, exceeds that of the observed one, $\Delta m_{\rm obs}$, we can estimate the probability of observing a microlensing event with amplitude larger than $\Delta m$. This corresponds to an empirical measurement of the p-value, which can be expressed as 
\begin{equation}
    p = P(\Delta m_{\rm sim}>\Delta m_{\rm obs})=\frac{N(\Delta m_{\rm sim} > \Delta m_{\rm obs})}{10^6}.
\label{eq:pvalues}
\end{equation}
We list the p-values ($p$) of the microlensing amplitudes in \tref{table:pvalues}. The p-values obtained by \cite{H20} are also provided for comparison. The difference between our results and the work of \citet{H20} is further discussed in \sref{sec:discussion}.
  
\begin{figure}
    \centering
        \includegraphics[width=\columnwidth]{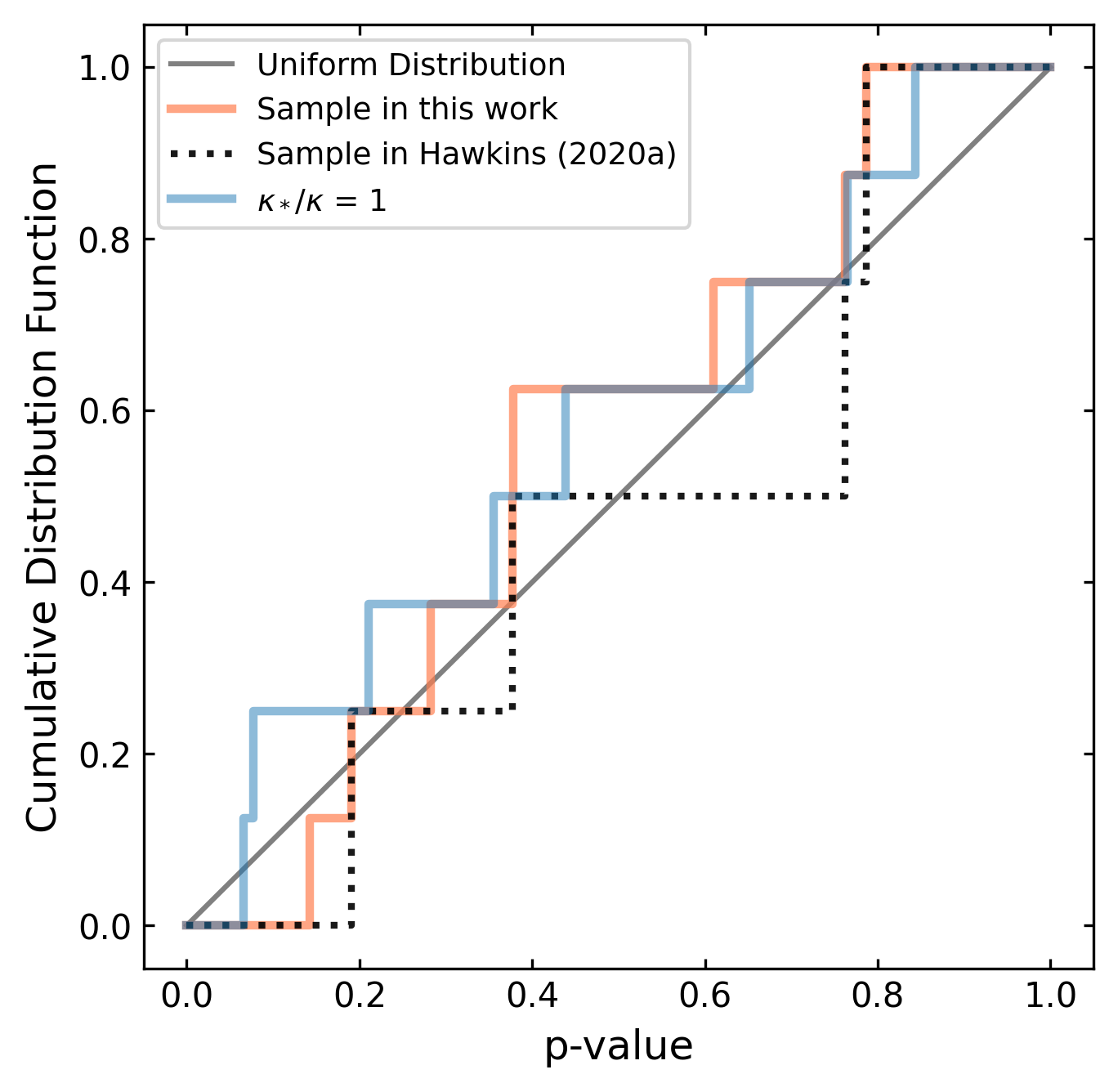}
    \caption{Cumulative distribution functions (CDF) of the p-values. The CDF of a uniform distribution represented by the solid grey line is what is expected when the hypothesis is true. Three tests represented by the dotted black, solid blue, and orange lines produce $p_{\rm KS}>0.05$, showing that the hypotheses of microlenses composed of compact objects only and of stars+smooth DM cannot be rejected due to the small sample size.}
    \label{fig:CDFCompare}
\end{figure}

\begin{figure*}
        \includegraphics[width=\textwidth]{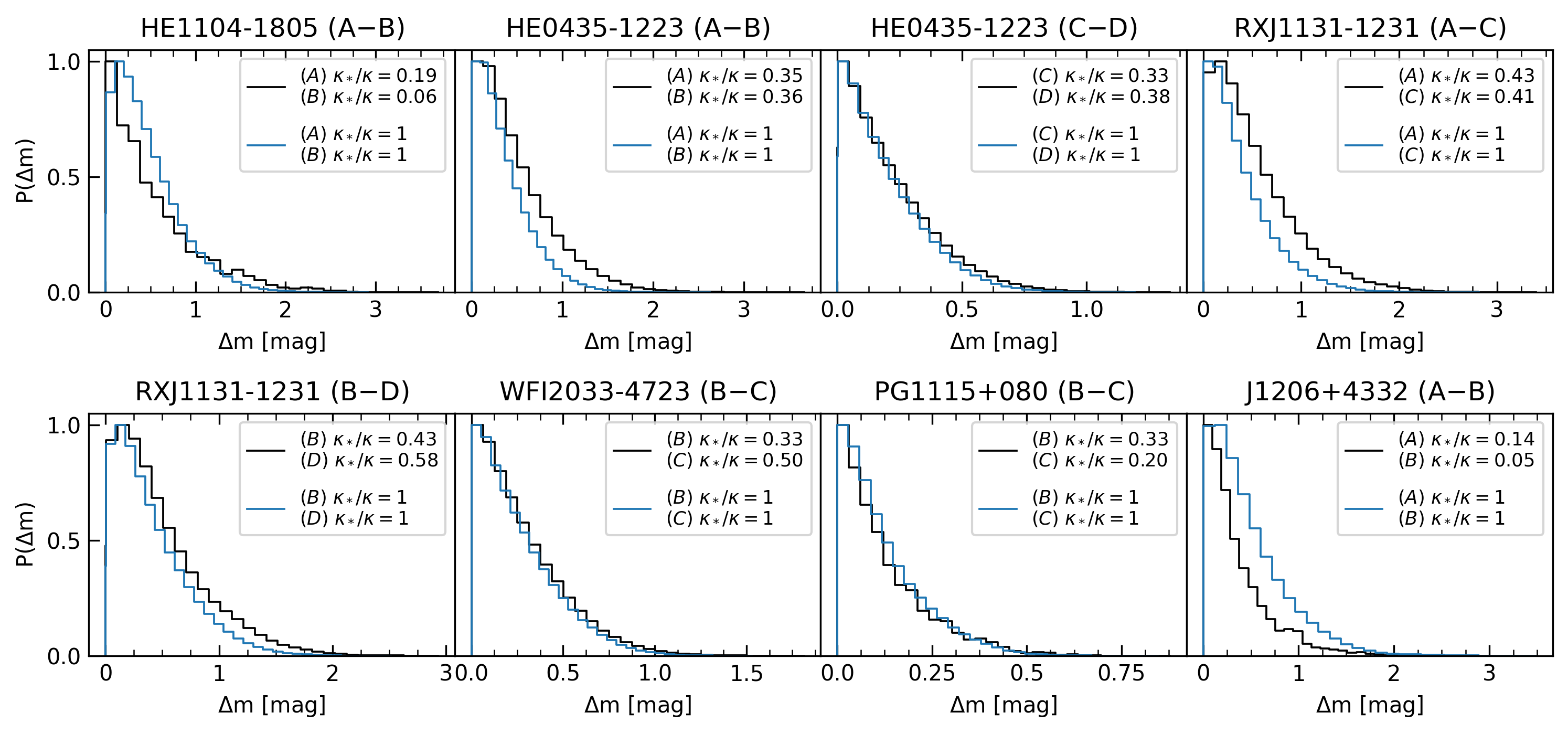}    
        \caption{Probability distributions of magnification amplitude, $\Delta m,$ using the LSST 10-year light curves, under the scenarios of stellar microlenses (black) and of stellar+DM microlenses (blue). The probabilities of two scenarios are employed to predict the number of difference light curves needed to distinguish between the two scenarios (see also \fref{fig:ksVSn}).}
    \label{fig:Discussion}
\end{figure*}

\section{Results}
\label{sec:results}

The probability distributions, $P(\Delta m)$, shown in \fref{fig:InitialModels}, lead us to the following remarks. For all systems, $\Delta m_{\rm obs}$ falls within one or two standard deviations around the median of the corresponding distribution. The largest difference from the mean can be seen in the case of RXJ1131, where $\Delta m_{\rm obs}$ is within two standard deviations from the median of the simulated distribution. This is because RXJ1131 is the system with the highest magnification amplitude (see \fref{fig:MicrolensingLC}). Such high magnifications are rare, but can still occur in  case of caustic crossings. This might well be the case of RXJ1131. 
Thus, even though the errors on the light curve measurements might be large in some cases (as seen in \fref{fig:MicrolensingLC}), there is no obvious indication that any of our observations for any individual object significantly deviate from expectations when assuming purely stellar microlensing.

We can go beyond the above phenomenological statements and quantify the agreement between our observations and microlensing model using statistical hypothesis testing. Traditionally, a popular choice is to use Pearson's $\chi^2$ test \citep{Snedecor&Coh} to compare expected outcomes with measurement distributions. However, the $\chi^2$ test requires a normality condition on the distributions we are comparing with and since our simulated amplitudes are not normally distributed as can be seen from \fref{fig:InitialModels}, a Pearson's $\chi^2$ test is not a viable option to use. 

Alternatively, we examine the uniformity of the p-value measurements to test our hypothesis using the Kolmogorov-Smirnov (KS) test  \citep{chakravarti1967handbook}. When a null hypothesis is true, the corresponding p-values appear to be distributed between 0 and 1 uniformly\footnote{The proof can be seen from \href{https://statproofbook.github.io/P/pval-h0}{https://statproofbook.github.io/P/pval-h0}}. Here, we define the null hypothesis $H_0$ as our conventional stellar microlenses model. Under this hypothesis, we test whether the probability distribution of $\Delta m_{\rm sim}$ matches the observational microlensing amplitude, $\Delta m_{\rm obs}$, using the p-values defined in \eref{eq:pvalues}. We then used the KS test to examine whether the eight p-values of our systems are indeed statistically compatible with a uniform probability distribution. Specifically, the KS test quantifies the distance between the cumulative distribution functions (CDFs) of the measured p-values and the uniform distribution. This distance can be converted to a single p-value $p_{\rm KS}$ (ranging from 0 to 1), showing how close these two CDFs are. We emphasize that $p_{\rm KS}$ is derived from the KS test, which is different from the p-value in \eref{eq:pvalues}. The resulting value for $p_{\rm KS}$ can be compared to any desired level of significance, $\alpha$. For example, if $p_{\rm KS}<\alpha=0.05$, we can conclude that our data rejects the null hypothesis $H_0$ at $95\%$ confidence level.

In \fref{fig:CDFCompare}, the orange line shows the CDF constructed for the p-values obtained in this work. We note that the CDF of the uniform probability distribution is the identity line $y=x$ (also shown in the figure). Performing the KS test to compare between these two CDFs yields $p_{\rm KS}=0.63 \gg \alpha=0.05$; we are therefore not able to reject $H_0$. We then display the sample  of \cite{H20} with the dotted line, which yields $p_{\rm KS}=0.87$. The null hypothesis in this case also cannot be rejected, which contradicts the finding in \cite{H20}. In \sref{sec:discussion}, we propose several possible reasons to explain such a difference.

The sample size in this work is  still small, even though it has been expanded by a factor of two compared to the sample in \cite{H20}. Supposing that we define an alternative hypothesis, $H_1$, stating that all DM behaves as compact objects, such as PBHs, we consider whether  our microlensing curves would then be able to reject $H_1$. Assuming that the mass distribution of PBHs follows the stellar distribution as a toy model, we generated the magnification maps by enforcing $\kappa_*=\kappa$, which means that all the mass of the galaxies is now in the form of compact objects with a mean mass of $0.2M_{\odot}$. We then repeated our analysis with this alternative model. The CDF of the new p-values resulting from this experiment is illustrated by the blue line in \fref{fig:CDFCompare}, yielding $p_{\rm KS}=0.89 \gg \alpha = 0.05$. Because of the small sample size, we conclude that it is unachievable to argue whether DM in a lensing galaxy is smooth or in compact form. However, the Rubin Observatory/LSST will provide thousands of lenses in the future, which may offer an opportunity to distinguish the hypotheses, $H_0$ and $H_1$, as presented in the next section.

\section{Predictions with ten years of LSST Data}
\label{sec:prediction}

The Rubin Observatory Legacy Survey of Space and Time \citep[LSST;][]{IvezicEtal19} will have an effective aperture of $6.4$-m with field of view of $9.6~\deg^2$. The LSST Camera will be devoted to a 10-yr imaging survey over $20\,000~\deg^2$. There will be $\approx3000$ lensed quasars (consisting of 2400 doubles and 600 quads) expected to be found in the LSST \citep{Oguri&Marshall10}, yielding $\approx3600$ independent difference microlensing light curves.

Given that the total convergence, $\kappa,$ of a lensed system is the sum of convergence created due to both the fraction of stars and DM (which could be smooth or compact as in PBHs), we test the ability of the upcoming observational data from the Rubin Observatory to discriminate between:
1) a fiducial scenario, where a percentage of the total convergence is in a form of compact bodies (such as stars), while the rest is still in a form of smooth matter; and 2) an alternative scenario: all the convergence is contributed by a form of compact bodies, namely, $\kappa_*/\kappa=1$.

In other words, in the fiducial scenario, $\kappa_*$ comes from the stellar contribution, while in the alternative scenario, we enforce $\kappa_*=\kappa$. The new set of light curves in the simulation is also extended to ten~years to match the duration of LSST measurements. The aim of this experiment is to estimate the number of light curves needed to discriminate between these two scenarios with a confidence level of 95\%. 

Following the simulations, the amplitude distributions for both scenarios are shown in \fref{fig:Discussion}, where the histograms in blue show the results for the alternative setup, while the histograms in black show the distributions for the fiducial setup where we used the macro-model parameters listed in \tref{table:MacroModels}. We note that the black histograms in \fsref{fig:InitialModels} and \ref{fig:Discussion} are slightly different due to the ten-year extension in the simulation.

\begin{figure}
    \centering
        \includegraphics[width=\columnwidth]{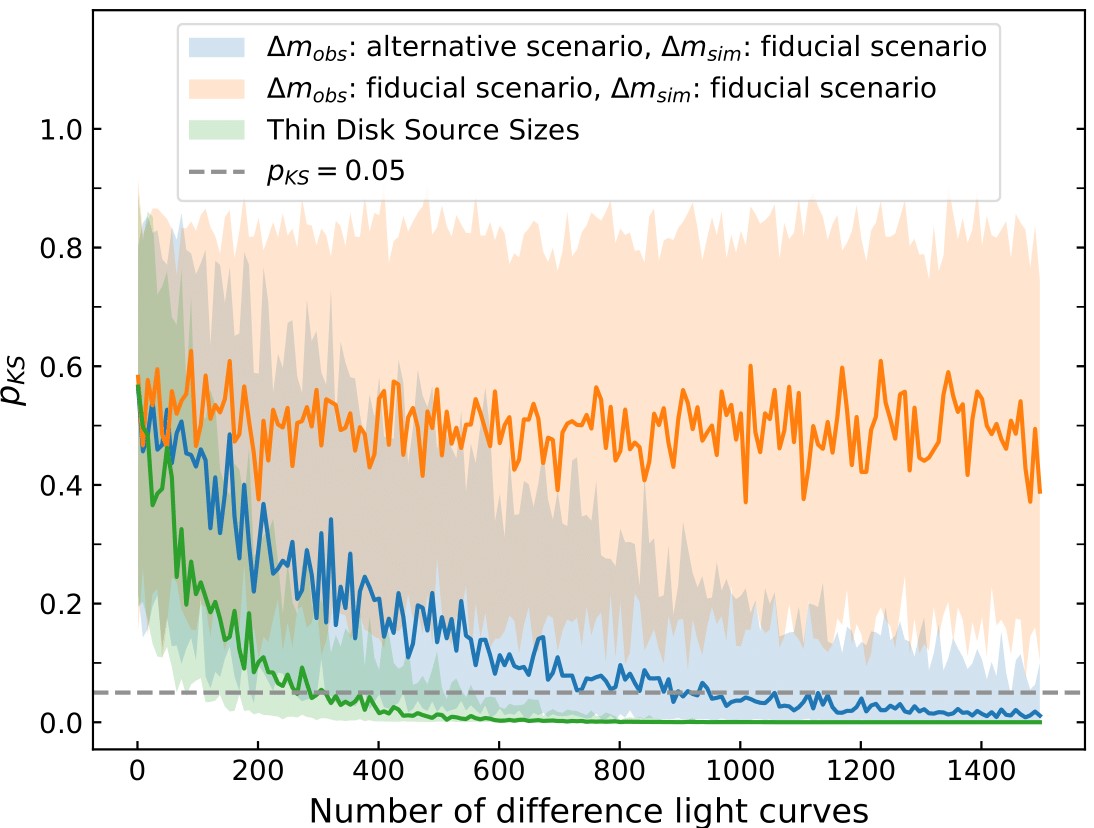}
    \caption{
    P-value from the KS test $p_{\rm KS}$ against number of microlensing light curves. The solid line indicate the median of $p_{\rm KS}$ distribution with the shade region bounded between the 84th and the 16th percentiles. The fiducial scenario adopts the conventional stellar distribution as microlenses, while the alternative scenario assumes that all the mass in the galaxy is in the form of compact objects, namely, $\kappa_*/\kappa=1$ (see \fref{fig:Discussion}). Given that $\Delta m_{\rm obs}$ and $\Delta m_{\rm sim}$ come from the same distribution, $p_{\rm KS}$ is expected to be 0.5 on average (orange line), regardless of the number of light curves used for this experiment. When $\Delta m_{\rm obs}$ and $\Delta m_{\rm sim}$ come from different distributions, the median value of $p_{\rm KS}$ decreases steadily with the sample size, enabling us to distinguish between two scenarios. The horizontal dashed line marks $\alpha=0.05$, corresponding to the threshold where the null hypothesis $H_0$ is rejected with a confidence level of $95\%$. The blue and green lines represent our results when using disk sizes from continuum reverberation mapping and from the thin disk model, respectively.}
    \label{fig:ksVSn}
\end{figure}

A microlensing light curve provides an amplitude of $\Delta m_{\rm obs}$. To mimic the observations, we randomly drew $\Delta m_{\rm obs}$ from the distribution in the alternative scenario ($\kappa_*=\kappa$) as a reference amplitude and calculated the p-value of its corresponding distribution in the fiducial scenario using \eref{eq:pvalues}. Once we obtained a set of p-values from a certain number of light curves, we performed the KS test to see if $p_{\rm KS}<\alpha$. When the number of light curves increases, $p_{\rm KS}$ decreases so that $H_0$ can be rejected with a confidence level of 95\% (as shown in the blue curve of \fref{fig:ksVSn}). Given a certain number of light curves, we further repeated this experiment 100 times in order to obtain the $p_{\rm KS}$ distribution, and its median is illustrated as the solid line with the shaded area bounded between the 16th and the 84th percentiles. The number of samplings of reference amplitudes is increased by picking more events randomly from the eight distributions of the COSMOGRAIL sample, since we have precisely measured lensing parameters ($\kappa$, $\gamma$, $\kappa_*$) only for these eight systems. Here, we make the assumption that the COSMOGRAIL sample is a representative sample of all LSST lensed quasars. We show that the median $p_{\rm KS}$ goes below $\alpha=0.05$ for about 900 curves. This means that it is possible to distinguish the fiducial and alternative scenarios with a confidence level of 95\%, provided that 900 image pairs are available. As we would expect, when sampling both $\Delta m_{\rm obs}$ and $\Delta m_{\rm sim}$ from the distributions in the fiducial scenario, the median $p_{\rm KS}$ oscillates around $0.5$, regardless of the number of light curves considered. This result is illustrated as the solid orange line in \fref{fig:ksVSn}, showing that $H_0$ can no longer to be rejected, since $\Delta m_{\rm obs}$ is drawn from the same distribution as $\Delta m_{\rm sim}$.

Given that the methodology of this work is dependent upon the source sizes (see the discussion in \sref{sec:discussion}), we re-performed our analysis using a different set of disk sizes for the chosen sample of lensed quasars, following the thin disk model \citep{MorganEtal10,Mosquera&Kochanek11}:   
\begin{equation}
    R_s = 9.7\times 10^{15}~{\rm cm}\ 
    \Big(\frac{\lambda_{\rm rest}}{\mu{\rm m}}\Big)^{4/3} \Big(\frac{M_{\rm BH}}{10^9 M_{\odot}}\Big)^{2/3} \Big(\frac{L}{\eta L_E}\Big)^{1/3},
\label{eq:ThinDisk}
\end{equation}
where the black hole mass of each system $M_{\rm BH}$ is obtained in Table~1 of \citet{Mosquera&Kochanek11}. We conventionally assume that the quasar accretion efficiency $\eta=0.1$ and the luminosity in units of the Eddington luminosity $L/L_E=1/3$ \citep{KollmeierEtal06,Hopkins&Hernquist09,Schulze&Wisotzki10}. The second set of sizes is then adopted as $R_{1/2}=2.44\ R_s$. We illustrate the results of the thin disk model as the solid green line in \fref{fig:ksVSn}. We can now observe that the number of curves needed to distinguish between the two scenarios drops to $\sim300$, as a result of the smaller estimates of the thin disk radii, which have the mean radius of $\sim4\times10^{15}$~cm ($\sim1.5$~light-days). The smaller the source size, the less blurred the caustic pattern. This then makes it easier to distinguish between the distribution of microlenses in either scenario. Given the span of disk size between 1.5 and 4 light-days, thus, the number of microlensing curves that are needed to test the validity of the compact DM scenario lies within a range from 300 to 900. Obtaining such a number of systems will be difficult but feasible given the capabilities of LSST. We thus propose, as an extension of this work, to apply this methodology to the future observations provided by LSST when the data becomes available.

\section{Discussion}
\label{sec:discussion}

In this section, we discuss the implications of several assumptions made in our work.  Concretely, we review the choices made for the source size and transverse velocity.
We also discuss the definition for the amplitude of the microlensing signals and the methodology developed for the calculation of their corresponding p-values.  

Since the convolution of magnification maps with a source brightness distribution smooths the caustic patterns, we expect the amplitude of microlensing variations to become smaller with increasing source size \citep{Refsdal&Stabell91,Witt&Mao94}. Therefore, we expect our results to vary as a function of the adopted disk sizes for our simulations, due to the uncertainties in the size measurements. In this work, we follow a physically motivated approach and we assume, for each system, its own source size with a half-light radius as calculated in the relevant literature. Although some microlensing studies can also provide disk size measurements \citep[e.g.][]{MorganEtal18,CornachioneEtal20}, we avoided using them, as they already included assumptions on the stellar microlenses in a halo of the lensing galaxy. Using their reported disk sizes might lead to the conclusion that stars are the only components needed to produce sufficient amplitudes in the lensed quasar microlensing curves, as this is the underlying hypothesis of their work. To overcome this circular argument, we  instead chose the disk sizes reported in recent continuum reverberation mapping studies, such as the work of \citet{MuddEtal18}, which is a technique that is independent of the assumption on the stellar populations (see \eref{eq:wavelengthScale}).We note that disk sizes from reverberation mapping are usually associated with large error bars. However, our method relies on a statistical treatment of the microlensing signal and not on individual measurements; thus, even though individual measurements may not be very precise, they serve well as an unbiased average measurement of the disk sizes. It is also possible to scale the source size with the black hole mass under the thin disk model (see \eref{eq:ThinDisk}), but since the thin disk size has frequently reported much smaller measurements than those from microlensing and reverberation mapping studies, we avoided choosing this scaling. 
As for our prediction in \sref{sec:prediction}, we present the results given two different sets of source sizes to highlight the possible range for the number of future microlensing curves needed to distinguish between the smooth DM regime and the compact DM regime.

In our analysis, we adopted $600$~km/s as a conventionally assumed transverse velocity for all systems. Indeed, any increase or decrease in this parameter value results in the respective lengthening/shortening of the trajectories drawn on the magnification maps, which could affect microlensing amplitudes in these simulated light curves. However, this bias is reduced with long-enough curves which are ensured by the long monitoring baseline of COSMOGRAIL. In our simulations, the trajectories are sufficiently long to explore a large span of dynamical range in the generated maps. Unlike the disk size, we therefore argue that the transverse velocity has marginal impact on our results.

This study is inspired by the works of \citet{H20,H20b}, who have suggested a need for other forms of compact objects to explain their measurements of the microlensing amplitudes. Our results, however, challenge these findings and agree with other works, such as \citet{EstebanEtal22}. We investigate a few possible causes to explain the inconsistency.
Firstly, a difference lies between the macro-model parameters used: while the lensing parameters from the TDCOSMO collaboration are considered to be more robust, the lensing models in \cite{H20} are obtained from a singular iso-thermal sphere plus an external shear model, which can produce different micro-caustic patterns and zero-points. Although \cite{H20} estimated the zero-points from the flux ratios of emission regions larger than the accretion disk, we notice that the measurements can deviate from those using lens modeling. 
Secondly, the metric for the microlensing amplitudes chosen in this work is less subject to biases from macro-magnifications, millilensing, dust absorption, and so on. Our empirical definition of $\Delta m$ does not require any measurement of the zero-points, which make them insensitive to this highly unreliable quantity. In order to understand how the lens models affect the hypothesis test, we repeated our analysis using the lensing parameters in Table~3 of \cite{H20} but using our definition of  $\Delta m$.  We still find that $H_0$ still cannot be rejected. 
Lastly, we think that the major difference between our work and that of \citet{H20} rests in the statistical treatments of the p-values when combining all considered systems. While we use the KS test to evaluate our stated hypothesis, the approach followed by \citet{H20} consists of simply taking the product of all calculated p-values. Since p-values are, by definition, smaller than 1, the latter statistical treatment will artificially disfavor the $H_0$ hypothesis as more and more systems are added to the sample. Our methodology avoids this biasing by following a statistical method for combining the information provided by each p-value.

Additionally, we discuss the effectiveness of our method in discriminating between the defined fiducial and the alternative scenarios, which is subject to the disk sizes of lens systems. This is explained by the fact that larger disk sizes yield smoother micro-caustic patterns on magnification maps, blurring the distinctions between the two compared scenarios. As a result, one would need more light-curves to reach a conclusion within $95\%$ confidence. In this work, we adopt the disk sizes from the reverberation mapping, and show that we need  -- at most -- about 900 microlensing curves to make the distinction between the two scenarios. We further explore the possibly smaller disks, scaled with the black hole masses using the typical thin disk model, which is about 2.5 times smaller than the reverberation mapping sizes. In this case, we only require $\sim300$ curves to test the validity of the compact DM scenario. Since accretion disks are smaller and more affected by microlensing at shorter wavelength, we emphasize that using bluer bands than the R band considered here will also sensibly increase the constraining power of this experiment. Although in this work, we test two extreme cases, it is possible to apply the same methodology to compare different scenarios under a broader range of assumptions; this exploration, however, will be left for a future work. Finally, LSST observations will be ongoing for total of $10$~years, consistently providing a greater number of realistic measurements of the microlensing amplitudes and, thus, helping set better constraints on the calculated p-values.

\section{Conclusion} 
\label{sec:conclusion}

In this work, we study the origin of high-magnification events in microlensing light curves of strongly lensed quasars. More precisely, we test whether stars in the lensing galaxies of a population of lensed quasars are sufficient to account for the observed microlensing signal. Our statistical analysis of the data leads to the following conclusions:

\begin{itemize}
    \item The considered population consists of the strongly-lensed quasar systems: HE~1104$-$1805, HE~0435$-$1223, RX~J1131$-$1231, WFI~2033$-$4723, PG~1115$+$080, and J0126$+$4332. From these lenses, eight independent pairs of lensed images are chosen and the microlensing curves for those pairs are used from the most recent light curves and time-delay measurements of COSMOGRAIL.
    
    \item We defined a robust metric to evaluate the amplitude of the microlensing signal and we computed, for each system, the p-value corresponding to the probability that our predicted microlensing amplitudes would be greater than the actual measurements.

    \item We proposed a coherent statistical approach to carry out a quantitative comparison of our microlensing observations and simulations. Our method utilizes the KS test to examine the uniformity of the distribution of p-values of our sample. This allows us to statistically test the null hypothesis, which posits that 1) galaxies are composed of stars and smoothly distributed DM against 2) an alternative scenario where all the mass of the galaxy is in the form of compact objects, namely, stars+PBH.
    
    \item On the basis of the current COSMOGRAIL sample, we demonstrated that our null hypothesis cannot be rejected statistically.
    
    \item We explored an alternative hypothesis in which DM is completely constituted by compact objects, such as PBHs, and also found that this hypothesis cannot be rejected. 
    
    \item Finally, we showed that $\sim 900$ microlensing curves are needed to ascertain the validity of either of the assumed scenario, within a $95\%$ confidence level. 
    
\end{itemize}

Our current sample of light curves from the COSMOGRAIL collaboration is the only and largest one to date poised to explore the possibility of using quasar microlensing to discriminate between different assumptions on the nature of microlenses. Although is is still too small, it lends considerable hope to the possibility of making the method truly effective, when hundreds of light curves become available from the Rubin Observatory/LSST in the coming years.


\section*{Acknowledgements}
We give our thanks to M.~Hawkins, D.~Sluse, G.~Vernardos, E.~Soutter, and M.~Auger for helpful discussions. This work is supported by the DSSC Doctoral Training Programme of the University of Groningen. This research was made possible by the generosity of Eric and Wendy Schmidt by recommendation of the Schmidt Futures program. This work is also supported by the European Research Council (ERC) under the European Union’s Horizon 2020 research and innovation program (COSMICLENS: grant agreement No 787886) and by the Swiss National Science Foundation (SNSF). M.M. acknowledges the support of the Swiss National Science Foundation (SNSF) under grant P500PT\_203114. 


\bibliographystyle{aa}
\bibliography{reference}



\end{document}